# Phonon and Thermal Properties of Exfoliated TaSe$_2$ Thin Films

Z. Yan[1], C. Jiang[1], T.R. Pope[2], C.F. Tsang[2], J.L. Stickney[2], P. Goli[3], J. Renteria[1], T.T. Salguero[2,*] and A.A. Balandin[1,3,*]

[1]Nano-Device Laboratory, Department of Electrical Engineering, Bourns College of Engineering, University of California – Riverside, Riverside, California 92521 USA

[2]Department of Chemistry, University of Georgia, Athens, Georgia 30602 USA

[3]Materials Science and Engineering Program, Bourns College of Engineering, University of California – Riverside, Riverside, California 92521 USA

## Abstract

We report on the phonon and thermal properties of thin films of tantalum diselenide (2H-TaSe$_2$) obtained via the "graphene-like" mechanical exfoliation of crystals grown by chemical vapor transport. The ratio of the intensities of the Raman peak from the Si substrate and the E$_{2g}$ peak of TaSe$_2$ presents a convenient metric for quantifying film thickness. The temperature coefficients for two main Raman peaks, A$_{1g}$ and E$_{2g}$, are -0.013 and -0.0097 cm$^{-1}$/°C, respectively. The Raman optothermal measurements indicate that the room temperature thermal conductivity in these films decreases from its bulk value of ~16 W/mK to ~9 W/mK in 45-nm thick films. The measurement of electrical resistivity of the field-effect devices with TaSe$_2$ channels indicates that heat conduction is dominated by acoustic phonons in these *van der Waals* films. The scaling of thermal conductivity with the film thickness suggests that the phonon scattering from the film boundaries is substantial despite the sharp interfaces of the mechanically cleaved samples. These results are important for understanding the thermal properties of thin films exfoliated from TaSe$_2$ and other metal dichalcogenides, as well as for evaluating self-heating effects in devices made from such materials.

*Corresponding authors: salguero@uga.edu (TTS) and balandin@ee.ucr.edu (AAB)

## I. Introduction

The successful exfoliation of graphene [1] and discoveries of its unique electrical [2, 3] and thermal [4-7] properties have motivated searches for other quasi two-dimensional materials with interesting properties, as well as techniques for their synthesis and characterization [8-12]. As a group, layered *van der Waals* materials [8] can be cleaved mechanically or exfoliated chemically by breaking the relatively weak bonding between the layers. The resulting thin films can be re-stacked into superlattice-type structures with various properties [8, 13, 14]. Some of us previously have studied quintuples—five atomic layers—of bismuth telluride ($Bi_2Te_3$), which reveal thermoelectric and topological insulator phenomena [13-16]. These quasi-2D crystals of $Bi_2Te_3$ have electrical and thermal properties that are substantially different from those of bulk $Bi_2Te_3$ [15, 16].

An interesting subgroup of inorganic van der Waals materials is the layered transition metal dichalcogenides ($MX_2$, where M=Mo, W, Nb, Ta or Ti and X=S, Se or Te) [17] Some of these materials manifest charge density wave (CDW) effects in the temperature range from ~100 K to room temperature [18]. In particular, we recently have found that decreasing the thickness of titanium diselenide ($TiSe_2$) thin films allows one to significantly increase the transition temperature ($T_P$) to the CDW phase [19]. This advance opens the possibility of CDW-based devices that can operate near or at room temperature.

In this paper we report on the physical properties of $TaSe_2$ films prepared by "graphene-like" mechanical exfoliation. We focus on their phonon and thermal properties, which are essential for the metrology of such films and for possible device applications. Indeed, apart from graphene and hexagonal boron nitride, layered van der Waals materials typically have rather low values of thermal conductivity, K [6, 13, 14]. Thin films of such materials should have even lower K values due to phonon–boundary scattering. However, because the thermal conductivity in such materials is strongly anisotropic (substantially larger in-plane) and the cleaved surfaces can be rather smooth, the change in K due to film thinning is difficult to predict. The degraded thermal conductivity can become a major impediment for practical applications due to unavoidable power dissipation and self-heating effects. The contribution of electrons to thermal transport in thin films of metal dichalcogenides is also not known precisely. Thus the main goals of this study are to characterize the phonon spectrum, the relative contributions of phonons and electrons to heat conduction and the extent of thermal conductivity degradation in $TaSe_2$ thin films.

The crystal structure of $TaSe_2$ consists of Se-Ta-Se layers containing Ta in trigonal prismatic coordination. The weak interlayer van der Waals bonding permits the facile mechanical delamination and exfoliation of such crystals, and it also leads to several $TaSe_2$ polytypes that differ in the relative orientation of the layers and their stacking arrangements. In this work, we specifically examine $2H$-$TaSe_2$. The structure of this polytype has a two-layer repeat pattern (AcA BcB), within which the trigonal prismatic units are rotated $60^o$ with respect to each other [20-22].

## II. Material Synthesis and Sample Preparation

The $2H$-$TaSe_2$ crystals used in this study were derived from $3R$-$TaSe_2$ powder according to standard methods [21]. Briefly, we first prepared $TaSe_2$ from the elements. After two firing cycles at 700 and 900 °C, the product consisted of mostly $3R$-$TaSe_2$ with traces of the $2H$-$TaSe_2$ form, which is typical for this synthetic route [21, 23]. Then this material was heated in the presence of $I_2$ within a sealed ampule with a 100 °C temperature gradient; these "chemical vapor transport" conditions [24] resulted in clusters of metallic gray crystals. As shown in Figure 1 (a, b, and d inset), these crystals have lateral dimensions of approximately 200-800 μm and thicknesses of approximately 5 to 30 μm. Powder X-ray diffraction (XRD) reveals that the major phase of crystallized material is the $2H$-$TaSe_2$ polytype; as shown in Figure 1c, the experimental diffraction pattern matches literature data for $2H$-$TaSe_2$ very well [25, 26]. In addition, we observe traces of $3R$-$TaSe_2$, in line with common observations of polytype mixtures within Ta-Se systems [20, 23]. The presence of this minor component does not affect subsequent samples that are derived from the mechanical delamination of $2H$-$TaSe_2$ single crystals.

Both energy-dispersive X-ray spectroscopy (EDS) and electron probe microanalysis (EPMA) data confirm that the composition of the crystals is within a few percent of the ideal stoichiometry (Figure 2 (d) and Table I). Several factors that may contribute to the observed variation include: (i) EPMA data were collected from a non-ideal crystal surface, which introduces additional error, (ii) incorporation of oxide impurities, and/or (iii) possible non-stoichiometry, which is known in the range of x<0.03 and 0.25<x<0.64 for 2H-$Ta_{1+x}Se_2$ [21, 23, 27]. In particular, X-ray photoelectron spectroscopy (XPS) indicates that the $TaSe_2$ crystal surfaces become oxidized upon exposure to air. Figures 1 (e) and (f) show XPS data for the Ta 4f and Se 3d photoelectric peaks from a series of samples. By comparison with a reference sample of Ta foil that exhibits both Ta metal and its native oxide $Ta_2O_5$ [28], we can see that 2H-$TaSe_2$ crystals also contain Ta oxide [29]. However, upon mechanical delamination of surface layers from the 2H-$TaSe_2$ crystals, the only remaining visible species is $TaSe_2$.

Thin films of 2H-$TaSe_2$ were prepared on Si/$SiO_2$ substrates following the standard "graphene-like" approach [1-3]. The thickness $H$ of the films ranged from a few trilayers to $H$=300 nm. The Se-Ta-Se atomic trilayer has a thickness $c$=6.359 Å and the lattice parameter $a$=3.434Å (see Figure 2). The crystal structure of 2H-$TaSe_2$ belongs to the space group $D_{6h4}$ and it has a unit cell consisting of two Se-Ta-Se trilayers. We used Raman spectroscopy as a metrology tool to verify the quality of the crystals and determine the thicknesses of the flakes. Raman spectroscopy (Renishaw InVia) was performed in the backscattering configuration under $\lambda$ = 633 nm laser excitation. An optical microscope (Leica) with a 50× objective was used to collect the scattered light. To avoid possible self-heating effects, the excitation laser power for Raman spectroscopic measurements was kept low at $P$<0.5 mW (on the sample surface).

### III. Raman Metrology of van der Waals Thin Films

Figure 3 (a) shows Raman spectra of $TaSe_2$ for nine exfoliated thin films with different thicknesses H ranging from a few nm to >250 nm. The thicknesses of the exfoliated films were measured by atomic force microscopy (AFM). The 2H-$TaSe_2$ crystal possesses 12 zone center lattice vibrational modes. Four of these modes ($E^2_{2g}$, $E_{1g}$, $E^1_{2g}$ and $A_{1g}$) are Raman active. The atomic displacements for these vibrational modes are shown in Figure 3 (b). The $A_{1g}$ and $E^1_{2g}$ modes can be clearly seen in the Raman spectra at ~235 $cm^{-1}$ and ~207 $cm^{-1}$, respectively (see Figure 3 (a)). The modes with the energy lower than 130 $cm^{-1}$ are blocked by the Rayleigh filter of the spectrometer. There have been few reports of Raman data for $TaSe_2$ [30-33]. In contrast, earlier studies used bulk $TaSe_2$ samples and focused on the lower frequency portion of the spectrum [30-32]. Comparison of our data with that in Ref. [33] indicates that the peak positions and mode assignments are consistent.

The pronounced Si peak from the substrate appears at 522 cm$^{-1}$. The intensity of the Brillouin zone-center Si peak is proportional to the interaction volume. It explains why the peak intensity is increasing with decreasing thickness of TaSe$_2$ film placed on top of Si substrate. The ratio of the intensity of Si peak to that of A$_{1g}$ or E$_{2g}$ can be used for determining the thickness of the exfoliated film. Figure 3 (b) presents the ratio of the intensity of the Si peak to that of the E$_{2g}$ peak, I(Si)/I(E$_{2g}$), as a function of thickness $H$. This ratio decreases exponentially with increasing film thickness and can be fitted with the equation $I(Si)/I(E_{2g}) = 11.2\exp(-H/54.9) - 0.11$ (here the thickness $H$ is in the units of nm). The obtained dependence can be used as a calibration curve for the Raman-based nanometrology of TaSe$_2$ films. The approach can be extended to other layered van der Waals materials. From a practical perspective, the Raman metrology of exfoliated films is easier and faster than AFM inspection.

The temperature dependence of Raman spectra of 2H-TaSe$_2$ were studied in the temperature range from 83 K to 613 K. The sample with exfoliated TaSe$_2$ flake was put in a cold-hot cell and the temperature can be controlled externally with 0.1 $^{\circ}$C accuracy. Figure 4 (a) shows the Raman spectrum of a typical TaSe$_2$ flake measured at different temperatures. The Raman peak positions of both A$_{1g}$ and E$^1_{2g}$ modes move to lower wavenumber range (red shift) when temperature increases. The Raman peak positions of TaSe$_2$ A$_{1g}$ and E$^1_{2g}$ modes at the different temperatures are shown in Figure 4 (b). In the measured temperature range, the temperature dependence of A$_{1g}$ and E$^1_{2g}$ modes can be represented by a linear relation $\omega = \omega_0 + \Gamma T$, where $\omega_0$ is the frequency of Raman peak when temperature T is extrapolated to 0 K, and $\Gamma$ is the first-order temperature coefficient. The extracted temperature coefficients of A$_{1g}$ mode and E$^1_{2g}$ mode are $\Gamma_1 = -0.013$ cm$^{-1}$/$^{\circ}$C and $\Gamma_2 = -0.0097$ cm$^{-1}$/$^{\circ}$C, respectively. The softening of A$_{1g}$ and E$_{2g}$ modes with increasing temperature is in agreement with the report for bulk 2H-TaSe$_2$ [30]. However, a direct quantitative comparison of the temperature coefficients is not possible because our data are for thin films rather than bulk and for a different temperature range.

## IV. Thermal Properties of TaSe$_2$ Thin Films

The obtained temperature coefficients characterize the inharmonicity of the TaSe$_2$ crystal lattice. They also can be used to extract the thermal conductivity data using the Raman optothermal technique, which was originally developed for graphene [5]. The data presented in Figure 4 (b) can be used as calibration curves for determining the local temperature rise in TaSe$_2$ flakes. The temperature is extracted from the shift of Raman peak positions. In a sense, the Raman spectrometer is thus used a thermometer. For the thermal measurements, we intentionally increase the intensity of the excitation laser so that it induces a local heating of the sample. The low thermal conductivity of TaSe$_2$ allows one to achieve local heating at the power level of about 1-2 mW. Figure 5 (a) shows a characteristic Raman mode of the tested TaSe$_2$ flake at low and high excitation power. The Raman peak position of E$_{2g}$ peak shifts from 209.1 cm$^{-1}$ to 207.7 cm$^{-1}$ as the power increase by 90%. The calculated change in the local temperature introduced by laser is ~ 144 $^{o}$C. The measurements were repeated using another Raman peak of TaSe$_2$ to ensure reproducibility of the local temperature measurement.

The temperature rise for a known dissipated power and sample geometry allow one to determine the thermal conductivity [5]. We have used the finite element method (FEM) to simulate the heat dissipation in the samples under test. The thermal conductivity was determined via the iteration approach. For the first simulation run, the thermal conductivity of TaSe$_2$ flake was assumed to have the value of K$_0$. The numerical simulation with the assumed value of thermal conductivity resulted in the modeled temperature rise $\Delta T_0$. The simulated temperature rise was compared with the experimental temperature rise $\Delta T$. If $\Delta T_0$ was larger (smaller) than $\Delta T$, the assumed value of the thermal conductivity K was increased (decreased) in the next run until the simulated temperature rise matched the experimental data and the final value of K was obtained. This procedure also requires the knowledge of the power dissipated in the thin film. The power level at the sample surface is measured directly with a detector. The absorption coefficient in the thin film can be deduced from the ratio of the integrated Raman intensity of Si peak from the substrate with TaSe$_2$ films placed on top of it to the Si peak of the substrate not covered with the TaSe$_2$ peak (see Figure 5 (b)). One can see that this ratio decreases when the film thickness increases and goes all the way to zero, which means that all laser power is absorbed in the film. The film thickness dependence of the Si peak intensity ratio can be fitted by an exponential decay function $I_{Si}(H)/I_{Si0} = 1.06\exp(-H/27.1) - 0.04$, where $H$ is the film thickness in unit of nm, and the extracted absorption coefficient, $\alpha$, is 27.1 nm$^{-1}$. The details of the FEM based procedure for extraction of thermal conductivity have been described in Ref. [34].

The inset to Figure 5 (b) presents a schematic of the simulated sample structure. The radius of simulated domain size is 100 μm and the radius of TaSe$_2$ flake and the laser spot are 4 μm and 0.5 μm, respectively. The thicknesses of TaSe$_2$ flakes used for extraction of thermal conductivity were H=45 nm, H=55 nm and H= 85 nm. The thickness of the silicon dioxide layer and silicon substrate were 300 nm and 0.5 mm, respectively. The silicon substrate was placed on an ideal heat sink so that the temperature at the bottom side was fixed at 300 K. The adiabatic conditions were assumed at other external boundaries. For the TaSe$_2$ flake with H=45 nm we obtained the experimental shift in the Raman peak position $\Delta\omega$=-1.4 cm$^{-1}$ as the power on the sample surface changed from 0.11 mW to 1.15 mW. This corresponds to a local temperature rise of $\Delta T$=144 K. The software-enhanced resolution of the micro-Raman spectrometer in these measurements was estimated to be 0.1 cm$^{-1}$, which sets the experimental error at about 10%. Errors related to radiative losses were estimated to be less than 10% due to the relatively fast time scale of the measurements. The results of the measurements are summarized in Table II. One can see that the room temperature thermal conductivity in these films decreases from its bulk value of about 16 W/mK to ~9 W/mK in 45-nm thick films. Although the electrical and thermal transport properties of graphite intercalation compounds and MX$_2$ materials were studied even earlier than mechanically exfoliated graphene [35-36], only a few prior reports have addressed the thermal properties of TaSe$_2$ and related materials [37-39]. It has been reported that the room temperature thermal conductivity of bulk 2H-TaSe$_2$ is ~16 W/mK [39]. Our results for mechanically exfoliated thin films are consistent with the prior measurements for bulk 2H-TaSe$_2$ [37-39].

For practical applications, it is important to understand whether the heat conduction is dominated by the phonon or electron contributions. To determine this distinction, we fabricated four-terminal devices with the channels implemented with thin film 2H-TaSe$_2$. We used electron beam lithography to define the top metal contacts for the electrical measurements. The top contacts were fabricated with 10 nm of Ti and 100 nm of Au. Figure 6 shows the current-voltage (I-V) characteristics of a representative device. The insets show the optical microscopy image of the 2H-TaSe$_2$ device and the linear region used for extraction of the electrical conductivity $\sigma$.

The channel lengths, widths and thicknesses were 9 µm, 6.5 µm and 80 nm, respectively. The experimentally determined conductivity of the 2H-TaSe$_2$ channel was $\sigma$=0.365 1/$\Omega$m for this device. Using the Wiedemann–Franz law, $K/\sigma T=(\pi^2/3)(k_B/e)^2$, we estimated that the contribution of electrons to thermal conductivity of thin films of intrinsic 2H-TaSe$_2$ is ~3x10$^{-6}$ W/mK. This means that the heat conduction is dominated by phonons.

Our data indicate that the thermal conductivity starts to decrease when the film thickness becomes comparable to the phonon mean free path in the material. The latter suggests that the observed decrease of the thermal conductivity, as compared to bulk value, is related to the acoustic phonon–boundary scattering. This is a non-trivial conclusion because one may have expected that in exfoliated thin films of van der Waals materials the smooth interfaces of the cleaved films will lead to negligible phonon – boundary scattering. It also is known that the thermal conductivity in MX$_2$ materials is strongly anisotropic (much larger in-plane K) and thus variations of thickness may not necessarily produce strong effects. Our results indicate that the thermal transport in exfoliated films on substrates is dominated by extrinsic effects, such as boundary scattering, rather than by the intrinsic lattice dynamic properties of van der Waals films.

## V. Conclusions

We investigated the phonon and thermal properties of 2H-TaSe$_2$ thin films prepared by the mechanical delamination of crystals. We have established that the intensities of Raman peaks from the films and the Si substrate can be used for the robust and fast nanometrology of TaSe$_2$ thin films. The Raman nanometrology method can be readily extended to other layered van der Waals materials. We also found that the thermal conductivity of the exfoliated thin films of 2H-TaSe$_2$ is dominated by the phonon contributions and reduced substantially compared to the bulk value. The Raman optothermal measurements indicate that the room temperature thermal conductivity in these films decreases from its bulk value of ~16 W/mK to ~9 W/mK in 45-nm thick films. These results suggest that thermal issues need to be taken into account in proposed device applications of TaSe$_2$, other metal dichalcogenide and van der Waals materials.

## METHODS

**Material Growth and Experimental Techniques:** For the TaSe$_2$ crystal growth, the elements were used as received: tantalum (99.9% trace metal basis) from Sigma-Aldrich Corp., selenium (99.99%) from Strem Chemicals Inc., iodine (99.9%) from J.T. Baker. For SEM and EDS, the samples were prepared by sprinkling crystals onto double-sided carbon tape. The images were acquired with an FEI Inspect F field emission gun scanning electron microscope operated at 20 keV. EDS data were collected with an EDAX Inc. instrument at 20 KeV for 19.7 s. For XRD, the samples were prepared by grinding the material with an agate mortar/pestle and then pressing into an aluminum mount. The patterns were acquired using a Bruker D8-Advance diffractometer (Co-K$\alpha$ radiation source) operated at 40 mA and 40 kV. Data were collected from 2-80 2$\theta$ at a rate of 0.1 s per step. For EPMA, the samples were prepared by placing crystals flat on double-sided carbon tape. The data were acquired on the crystal surfaces parallel to the substrate using a JEOL JXA 8600 Superprobe. XPS analyses were performed using a Mg K$_{\alpha1,2}$ (STAIB) source and a hemispherical analyzer from Leybold Heraeus; the take-off angle was about 40°. The preparation of TaSe$_2$ followed the following steps. The stoichiometric amounts of elemental Ta and Se were sealed in a well-evacuated quartz ampule (backfilled 3x with high purity Ar). This mixture was heated at 700 °C for 15 hours, cooled to room temperature, and then thoroughly mixed by shaking. This material was heated again at 900 °C for an additional 15 hours. The final product was a black, glittery, free-flowing powder. The chemical vapor transport of TaSe$_2$ used the following procedures. The previously synthesized 3R-TaSe$_2$ was loaded into a quartz ampule with 3 mg cm$^{-3}$ of I$_2$. The ampule was evacuated (while immersed in a dry ice/acetone bath) and sealed, then heated with a ramp of 1 °C min$^{-1}$ in a tube furnace until a gradient of 730 to 830 °C (from the cool end to the hot end of the ampule) was achieved. Those temperatures were held for one week, then the sample was cooled to room temperature at 1 °C min$^{-1}$. After removal from the ampule, the metallic grey crystals were warmed slightly under vacuum to remove any traces of I$_2$.

**Numerical Modeling Approach for Thermal Data:** The modeling of heat diffusion in TaSe$_2$ thin films on Si substrate was performed using the finite element method implemented in COMSOL software package with the Heat Transfer Module. The schematic of the simulated sample structure is presented in the inset to Figure 5 (b). The substrate consists of 0.5 mm Si and 300 nm SiO$_2$ layer. The radius of the Si/SiO$_2$ used in the simulated domain is 100 µm. The thickness and radius of TaSe$_2$ layer varied for different flakes according to the values obtained experimentally. The radius of laser spot size is 0.5 µm. The heat conduction was simulated by solving numerically Fourier's law $-\nabla \cdot (K \nabla T) = Q$, where $K$ is the thermal conductivity, $T$ is the absolute temperature and $Q$ is the power density of the heat source. The thermal conductivity values of Si and SiO$_2$ used in the simulation were 140 W/mK and 1.4 W/mK, respectively. The absorbed laser power density along z direction can be described by the following function $Q(z) = Q_0 \cdot \alpha \cdot \exp(-\alpha \cdot z)$, where $Q_0$ is the laser power density at the top surface of the sample, $\alpha$ is the absorption coefficient of TaSe$_2$ film obtained in previous steps. For a given dissipation power and sample geometry, the numerical modeling generates a temperature profile in the sample structure. Due to the axial symmetry of the simulation domain around z direction, the temperature profile in the whole sample structure can be fully represented by the simulation results in z-ρ cross-section, where r is coordinate in the cylindrical coordinate system. The thermal conductivity of thin films of TaSe$_2$ films was determined via the iteration approach as described in Ref. [34].

*Acknowledgements*

This work was funded, in part, by the National Science Foundation (NSF) and Semiconductor Research Corporation (SRC) Nanoelectronic Research Initiative (NRI) project 2204.001: Charge-Density-Wave Computational Fabric: New State Variables and Alternative Material Implementation (NSF-1124733) as a part of the Nanoelectronics for 2020 and beyond (NEB-2020) program. AAB also acknowledges particle support from the STARnet Center for Function Accelerated nanoMaterial Engineering (FAME) – SRC program sponsored by MARCO and DARPA.

**Table I: Comparison of EDS and EPMA data from 2H-TaSe$_2$ crystals**

|  | Theoretical Weight % | Experimental Weight % | | Theoretical Atomic % | Experimental Atomic % | |
|---|---|---|---|---|---|---|
|  |  | EPMA | EDS |  | EPMA | EDS |
| Se | 46.60 | 47.60 | 43.23 | 66.00 | 67.80 | 63.57 |
| Ta | 53.40 | 51.86 | 56.77 | 33.00 | 32.24 | 36.43 |
| Total Weight % | 100.0 | 99.46 | 100.0 | --- | --- | --- |
| Se/Ta | --- | --- | --- | 2.00 | 2.10 | 1.74 |

**Table II: Thermal conductivity of exfoliated 2H-TaSe$_2$ thin films at room temperature**

| Thickness H (nm) | 45 | 55 | 85 | Bulk |
|---|---|---|---|---|
| K (W/mK) | 9 | 11 | 16 | 16 |

**FIGURE CAPTIONS**

**Figure 1:** Characterization of 2H-TaSe$_2$ crystals grown by the chemical vapor transport method: (a) Optical microscopy image (scale bar is 250 μm). (b) Scanning electron microscopy (SEM) image (scale bar is 30 μm). (c) Powder X-ray diffraction data. The top red trace shows the pattern from as-grown TaSe$_2$ crystals; the primary component is 2H-TaSe$_2$ (database comparison shown at bottom) with traces of 3R-TaSe$_2$ (marked with *). (d) Energy dispersive X-ray spectroscopy data for a 2H-TaSe$_2$ crystal (SEM inset, scale bar is 100 μm). (e) X-ray photoelectron spectroscopy (XPS) spectra showing the Ta 4f photoelectric peaks from 2H-TaSe$_2$ crystals before (red curve) and after (blue curve) peel-off of surface layers. The spectrum of Ta$_2$O$_5$ on Ta foil (black curve) is included for reference. (f) XPS spectra showing the Se 3d photoelectric peaks from 2H-TaSe$_2$ crystals before (red curve) and after (blue curve) peel-off of surface layers.

**Figure 2:** Crystal structure of 2H-TaSe$_2$ (a) and schematic of its main vibrational modes.

**Figure 3:** (a) Raman spectrum of the exfoliated thin films of TaSe$_2$ on Si substrate. The data is shown for the films thickness ranging from ~260 nm to below 30 nm. The characteristics E$_{2g}$ and A$_{1g}$ peaks of TaSe$_2$ are clearly observed. The intensity of Si peak at 522 cm$^{-1}$ is increasing with the decreasing thickness of TaSe$_2$ film. (b) The ratio of the intensity of Si peak to E$_{2g}$ peak in Raman spectrum of TaSe$_2$. The calibrated intensity ratio can be used for nanometrology of TaSe$_2$ films. The insets show optical microscopy images of two flakes of TaSe$_2$ with substantially different thickness. The thinner flakes appear blue in color whereas the thicker flakes appear yellow.

**Figure 4:** (a) Evolution of Raman spectrum of 2H-TaSe$_2$ thin film with temperature. A new peak appears in the spectra at about 250 cm$^{-1}$ after temperature decreases below ~220 K. The peaks shift their position with temperature. (b) Temperature coefficients Γ$_1$ and Γ$_2$ for E$_{2g}$ and A$_{1g}$ Raman peaks of 2H-TaSe$_2$, respectively. The temperature coefficients can be used for the extraction of the thermal conductivity of thin films using the Raman optothermal method.

**Figure 5:** (a) Shift and broadening of the Raman peaks of 2H-TaSe$_2$ as a result of local heating with the excitation laser. (b) Intensity ratio of the Si Raman peak from the substrate without thin

film to that of the substrate covered with 2H-TaSe$_2$ thin film. The data allows one to estimate the amount of power absorbed by the film. The inset shows a model used for extraction of the thermal conductivity of thin films of 2H-TaSe$_2$.

**Figure 6:** Current-voltage characteristic of a two-terminal device with 2H-TaSe$_2$ channel. The upper inset show optical microscopy image of the device. The lower inset shows the small-bias region of I-V curve used for determining the resistivity of the exfoliated flake of 2H-TaSe$_2$.

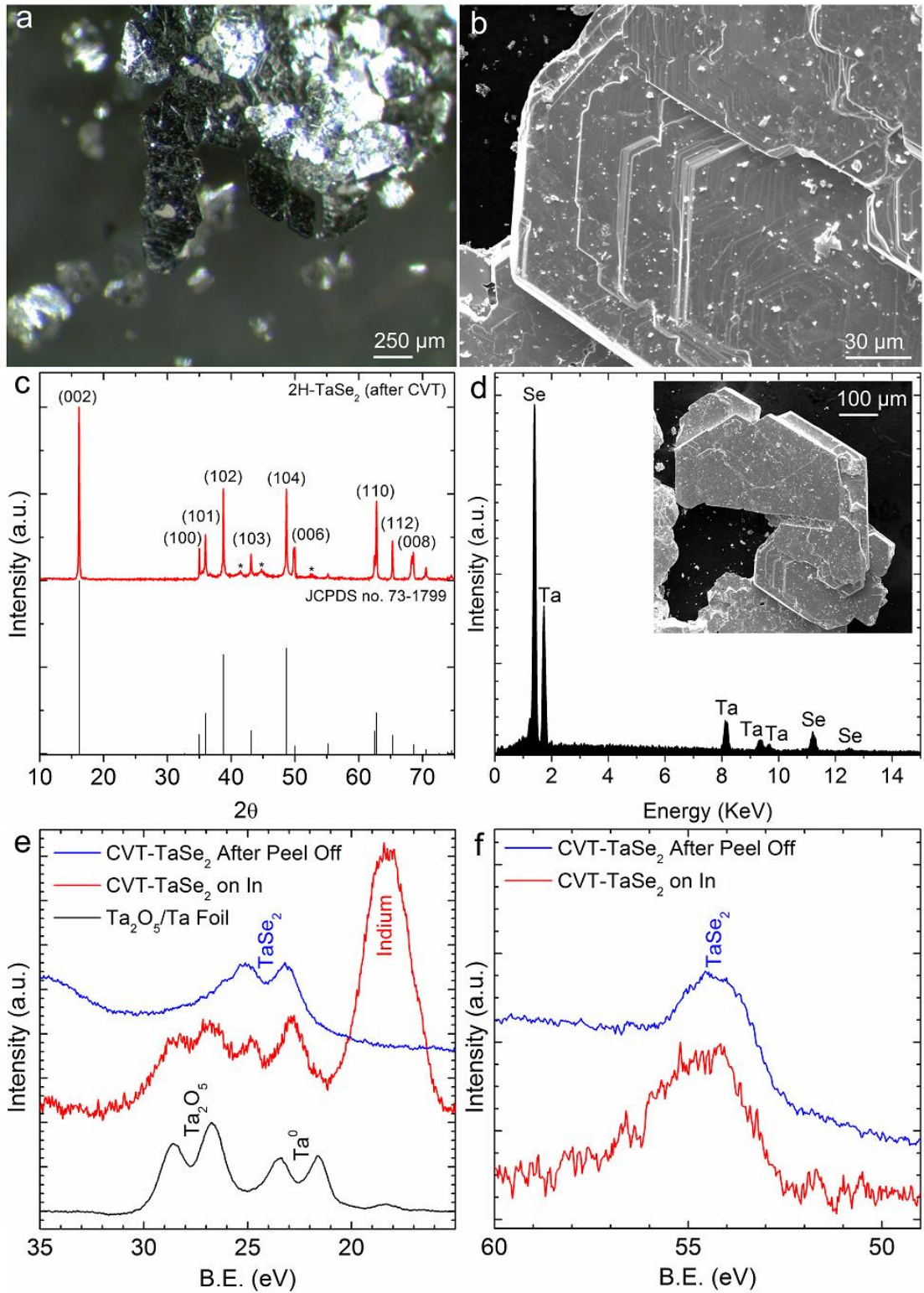

Figure 1

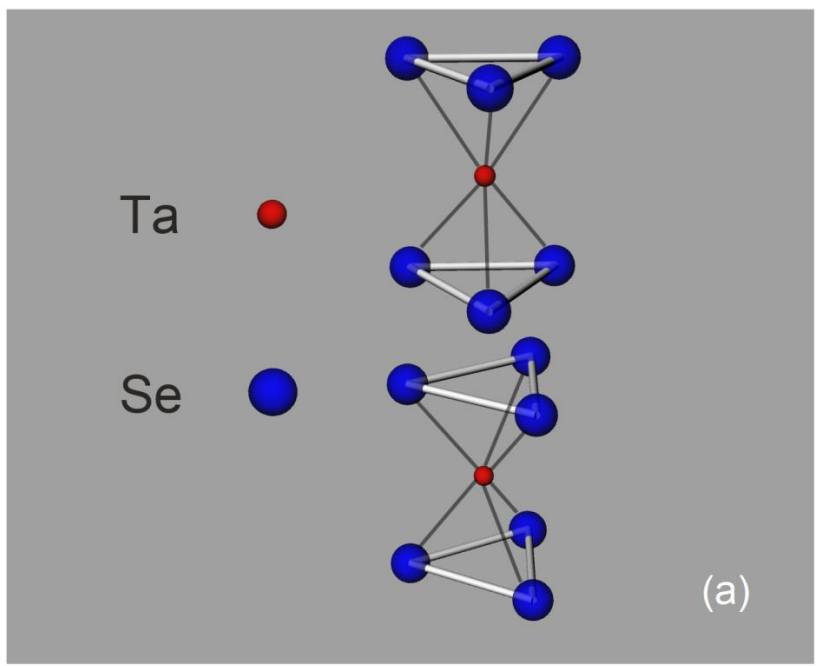

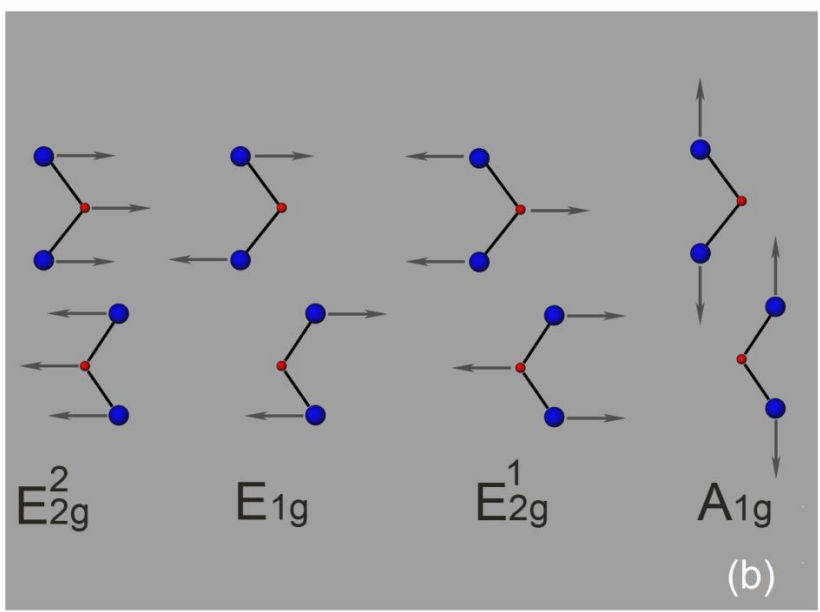

Figure 2

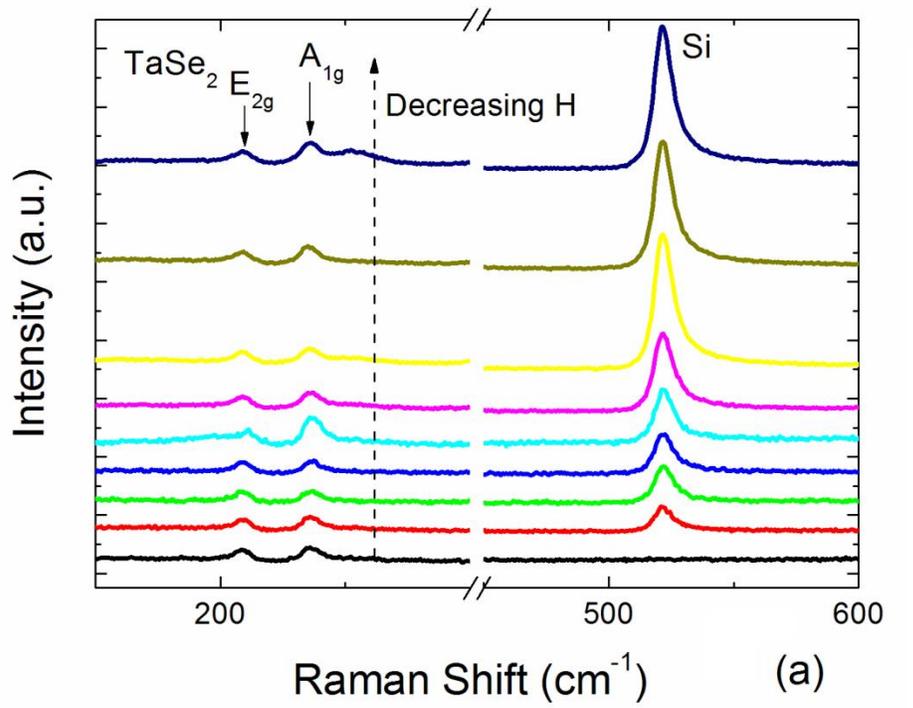

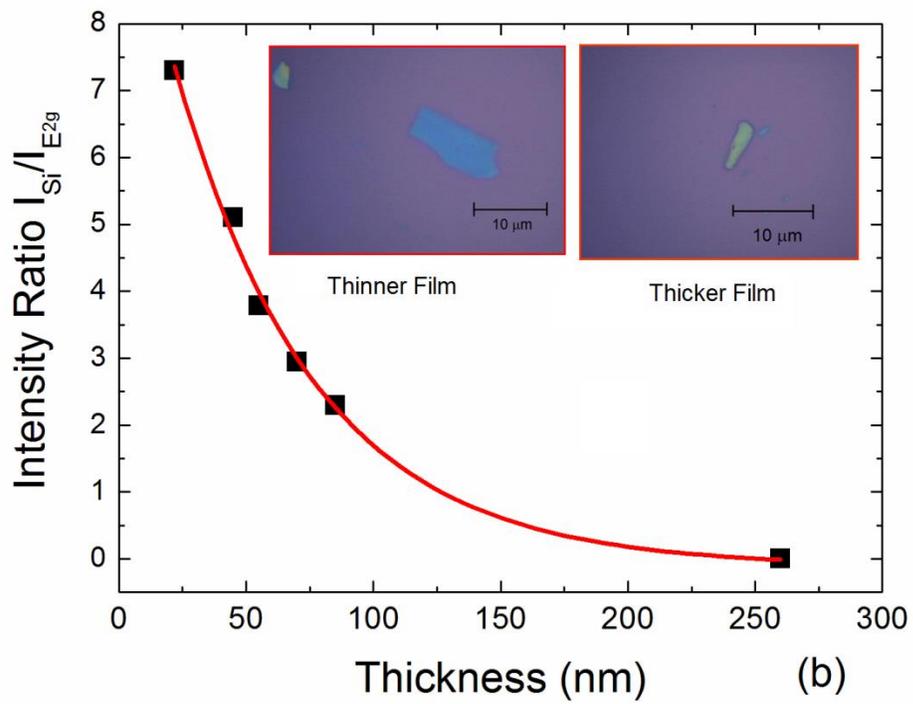

Figure 3

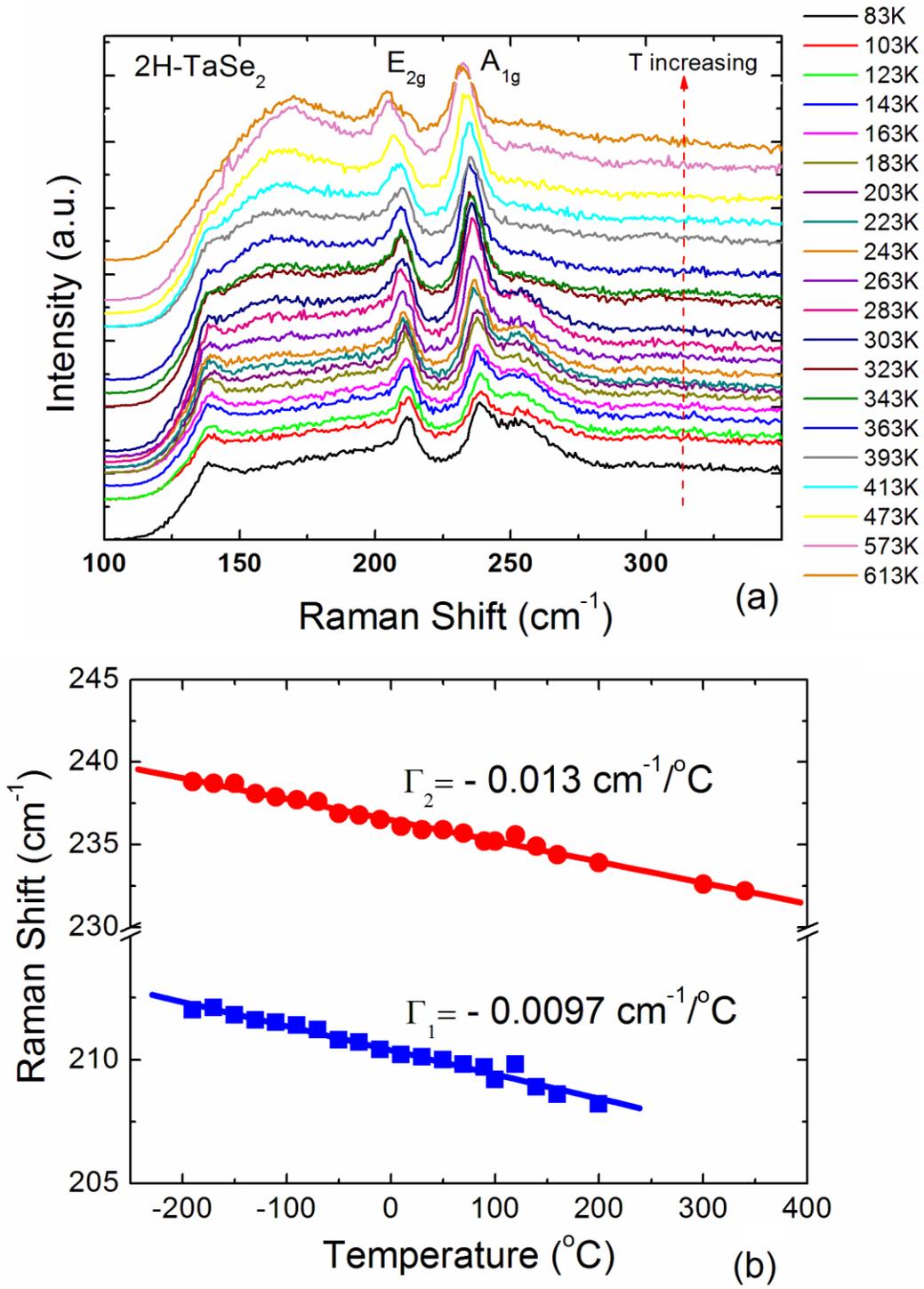

Figure 4

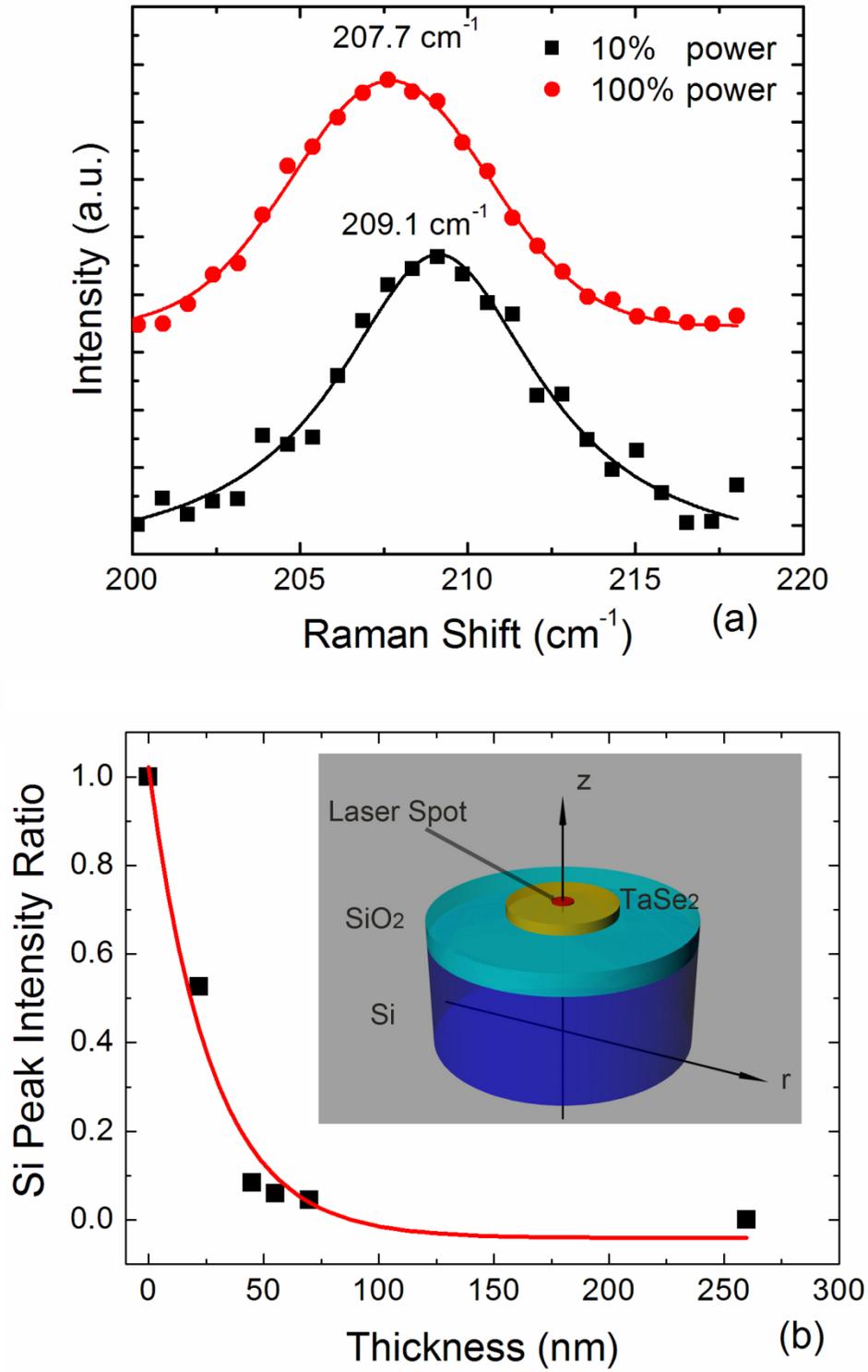

Figure 5

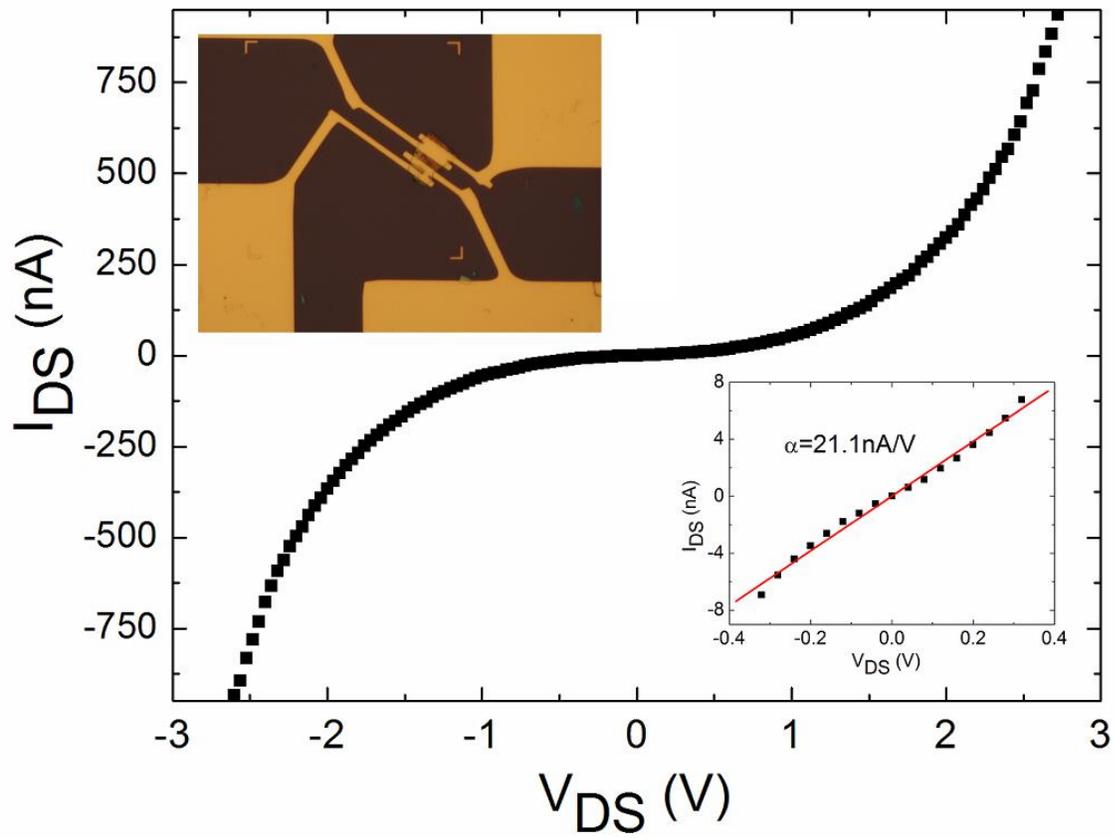

Figure 6